\documentclass{rcdj}

\begin{document}
\setcounter{page}{43}

\title{ON THE HISTORY OF THE DEVELOPMENT\\ OF THE NONHOLONOMIC DYNAMICS}
\runningtitle{ON THE HISTORY OF THE DEVELOPMENT OF THE NONHOLONOMIC DYNAMICS}
\runningauthor{A.\,V.\,BORISOV, I.\,S.\,MAMAEV}
\authors{A.\,V.\,BORISOV}
{Department of Theoretical Mechanics\\
Moscow State University,
Vorob'ievy Gory\\
119899, Moscow, Russia\\
E-mail: borisov@rcd.ru}
\authors{I.\,S.\,MAMAEV}
{Laboratory of Dynamical Chaos and Nonlinearity\\
Udmurt State University, Universitetskaya, 1\\
426034, Izhevsk, Russia\\
E-mail: mamaev@rcd.ru}
\abstract{}
\journal{REGULAR AND CHAOTIC DYNAMICS, V.\,7, \No1, 2002}
\received 09.11.2001.
\amsmsc{70E18, 70E40}
\doi{10.1070/RD2002v007n01ABEH000194}
\maketitle

\vspace{-1cm}

It is possible to single out two directions in the development of the nonholonomic
dynamics, and in each of these directions one can find interesting
investigations. One of these investigations is connected with the general
formalism of the equations of dynamics that differs from the Lagrangian and
Hamiltonian method of the equations of motion's construction.
Historically, certain errors of the well-known mathematicians, among which
we can name C.\,Neumann~\cite{Neuman} and E.\,Lindel\"of~\cite{lindelef},
were caused by an incorrect application of the Lagrange equations in the presence
of nonintegrable constraints in the description of the problem of a body rolling
without sliding on the horizontal plane. The general understanding of
inapplicability of Lagrange equations and variational principles to the
nonholonomic mechanics is due to H.\,Herz, who expressed it in his
fundamental work \emph{Die Prinzipien der Mechanik in neuem
Zusammenhange dargestellt}~\cite{hertz} that deals mostly with his conception of hidden cyclic
parameters (coordinates, masses), as opposed to the conventional notion of
interaction as a result of force application.

Herz's observations were developed by H.\,Poincar\'e~\cite{poin-gertz} in
his well-known paper \emph{Herz's Ideas in Mechanics}. From this paper, we
would like to quote some paragraphs, which illustrate, though in somewhat
epistolary form, the views of both scholars.\medskip

{\small ``Herz terms a system \emph{holonomic} when the following holds:
if the system's constraints do not allow a direct transition from one
position to another infinitely close position, then they either do not
allow indirect transitions between these positions. Only rigid constraints
exist in such systems.

It is evident that our sphere is not a holonomic system.

So, it sometimes happens that the principle of least action cannot be
applied to nonholonomic systems. Indeed, one can proceed from position~$A$
to position~$B$ taking the path that we have just discussed, or,
undoubtedly, one of many other paths. Among these, there is, evidently,
one path corresponding to the least action. Hence, it should have been
possible for the sphere to follow this path from~$A$ to~$B$. But this is
not so: whatever the initial conditions of motion may be, the sphere will
never pass from~$A$ to~$B$.

In fact, if the sphere does pass from position~$A$ to position~$A'$, it
does not always follow the path that corresponds to the least action.

The principle of least action holds no more.

Herz says, ``In this case, a sphere obeying this principle would seem to
be a living creature, which deliberately pursues a certain goal, while a
sphere following the laws of Nature would look as an inanimate
monotonously rolling mass\dots\ But such constraints do not exist in
Nature. So-called \emph{rolling without sliding} is actually rolling with
slight sliding. This phenomenon belongs to the class of irreversible
phenomena, such as friction; these phenomena are still poorly
investigated, and we have not yet learned to apply to them the true
principles of Mechanics.''

Our reply is, ``Rolling without sliding does not contradict either the law
of energy conservation, or any other law of physics known to us. This
phenomenon can be realized in the observable world within the accuracy
that would allow its application to construction of the most accurate
integration machines (planimeters, harmonic analyzers, etc.). We have no
right to exclude it from consideration as impossible. As for our problems,
they still remain regardless of whether such rolling is realized exactly
or only approximately. To accept the principle, it is necessary to require
that its application to a problem with almost exact source data would
yield the results, as close to the exactness as the source data were.
Besides, other (rigid) constraints can also be realized in Nature only
approximately. But nobody is going to exclude them from
consideration\dots''

}\medskip

The basic difference between nonholonomic dynamics and common Lagrangian
one lies in the fact that the equations of constraints, written in terms
of generalized coordinates~$q_j$ and generalized velocities~$\dot q_j$ in
the following form:
\begin{equation}
\label{intro-1} f_i(\bq,\dot\bq,t)=0,\qquad i=1\dts k, \quad \bq= (q_1\dts
q_n),
\end{equation}
cannot be presented in the final (integral) form
\begin{equation}
\label{intro-2} F_i(\bq,t)=0
\end{equation}
that sets limits only to the generalized coordinates. In this sense, one
says that the constraints are nonintegrable (differential). According to
Herz~\cite{hertz}, they also can be called nonholonomic\footnote{The term
\emph{holonomic} was coined from two Greek roots $\ddot o\lambda o\zeta$
(whole, integrable) and $\nu o \mu o\zeta$ (law).}.

Historically, Ferrers's equations with undetermined
multipliers~$\lambda_1,\ldots,\lambda_k$ (1871)
\begin{equation}
\label{eq-pref-1}
\frac{d}{dt}\Bigl(\pt{T}{\dot
q_i}\Bigr)-\pt{T}{q_i}=Q_i+\suml_j\lambda_j\pt{f_j}{q_i}.
\end{equation}
should be considered as the first general form for the equations of
nonholonomic mechanics. In equations~(\ref{eq-pref-1}), $T$ is the kinetic
energy, $Q_i$ stand for the generalized forces, while $\lambda_j$ are
undetermined multipliers, which, generally speaking, can be unambiguously
recovered from the constraint condition $f(q,\,\dot q)=0$. As a rule, the
constraints studied in nonholonomic mechanics are linear with respect to the generalized
velocities, i.\,e. realizable in conceptual problems,
\begin{equation}
\label{svyasi} f_i(\bq,\dot{\bq},t)=\suml_k a_{ik}(\bq,t)\dot
q_k+a_i(\bq,t)=0.
\end{equation}
However, Bolzmann and Hamel refer to a somewhat artificial example of a
nonlinear nonholonomic constraint. Ferrers also excluded the undetermined
multipliers and obtained a sort of an analog of Lagrangian equations of
motion~\cite{ferrers}.

Besides Ferrers's equations, nonholonomic dynamics also makes use of the
equations that are due to Appell, Chaplygin, Magri, Volterra, and Bolzmann\f
Hamel. Such a diversity, generally speaking, can hardly be regarded as a
major achievement. All these forms result from various ways of excluding
the undetermined multipliers, and usually are not of much practical use.
When setting up the specific equations to describe, for example, the
process of rolling, one would usually use the general dynamic equations or
the original equations in form~\eqref{eq-pref-1}.

The second direction substantially more important, in our opinion, for dynamics
in general, this direction includes investigations concerning the analysis of
the specific nonholonomic problems. The early statements of such problems
date back to E.\,Routh, S.\,A.\,Chaplygin, P.\,V.\,Woronetz, P.\,Appell, and
G.\,K.\,Suslov; these people have found remarkable integrable situations
and provided them with proper analytical and qualitative descriptions. The
majority of these problems also deal with rolling bodies. Together with
searching for integrable cases, many researches were concerned with
the stability of particular solutions (permanent rotations, for example) of the
general, nonintegrable, systems. Most widely known are, for example, the
investigations concerning the stability of rotations of so-called \emph{Woblestone}
about its vertical axis; this stone shows a surprising dependency
of its stability on the direction of its rotation. The most complete
analytical results for this phenomenon were obtained by
A.\,V.\,Karapetyan, I.\,S.\,Astapov, A.\,P.\,Markeev, and M.\,Pascal, but
their work nowise exhausted the problem of description of the evolution of
this system (some preliminary numerical results can be found
in~\cite{lindberg}). Certain properties of Woblestone with its
distinctive noncoincidence between the geometrical and dynamical axes are
still waiting for the proper theoretical explanation.

In the recent two decades, the development of the nonholonomic systems'
studies has been associated with finding of new integrable problems by
V.\,V.\,Kozlov, A.\,P.\,Markeev, A.\,P.\,Veselov, L.\,E.\,Veselova,
Yu.\,N.\,Fedorov, A.\,V.\,Borisov, I.\,S.\,Mamaev, as well as with
computer-aided and qualitative investigations of integrable and
nonintegrable situations. In this connection, one should mention
papers~\cite{cushman2,hermans,kholm1,kholm2,oreilly}. The majority of the
said results open new horizons in the study of nonholonomic systems.

On the other hand, we would like to draw reader's attention to some recent papers
dealing with the problems of nonholonomic reduction and almost Hamiltonian
formulation of the equations of nonholonomic
mechanics~\cite{bates,bates1,bloch,cushman}. These papers develop the
method of reduction with the help of the symmetry, which, even for the case of Hamiltonian
situation, has been over formalized by Marsden and Weinstein, rendering
even the simplest facts dating back to Jacobi and Poisson completely
unevident. Again, this formalism does not help to solve any new
problem, neither its development in the nonholonomic case gives any
conceptual results. As for the formulation of the equations of motion in
the almost Hamiltonian form~\cite{vanderschaft}, for which Jacobi identity
does not hold, one can only assert that such formulation in itself has
only formal meaning, though certain dynamical effects exist that prevent
the equations of nonholonomic mechanics from being formulated in the
``true'' Hamiltonian form. One of these effects, associated with the
nonexistence of the invariant measure and with asymptotic properties, was
mentioned by V.\,V.\,Kozlov in his fundamental work~\cite{kozlov1}.

Unfortunately, the papers by
S.\,A.\,Chaplygin~\cite{sani,chaplygin,chaplygin1}, G.\,K.\,Suslov~\cite{suslov},
Wagner~\cite{vagner}, V.\,V.\,Kozlov were never translated into English,
thus remaining unknown to the majority of the world's scientific
community. The publication of the present and the next
volumes is meant to partially fill the vacuum.

Let us also mention comparatively new and quite remarkable studies by
V.\,A.\,Yaroshchuk~\cite{yar,yar1}, who found new cases of the invariant
measure's existence, as well as papers by
A.\,V.\,Karapetyan~\cite{karapet} and V.\,V.\,Kozlov~\cite{kozlov3},
concerning the question of realization of nonholonomic constraints. These
papers develop the earlier studies by C.\,Caratheodori~\cite{carath}, who
associated the question of the origin of nonintegrable constraints with
infinitely large viscosity factor. To the readers interested in other
models of the dynamics of systems with nonintegrable constraints, Dirac
mechanics and vaconomic mechanics, we recommend
reviews~\cite{dirak,weber}.

The modern achievements in the studies of stability of nonholonomic
systems are described in~\cite{karapet2}, more elementary issues are
discussed in~\cite{neimark,ruban}.

There are only two special monographs dealing with nonholonomic mechanics:
by Yu.\,I.\,Neimark, N.\,A.\,Fufaev~\cite{neimark} and by
V.\,V.\,Dobronravov~\cite{dobro}. Both are quite out-of-date, and the
latter contains a number of incorrect assertions. In newer reviews by
P.\,Griffiths~\cite{griffits}, A.\,M.\,Vershik and
V.\,Ya.\,Gershkovich~\cite{vershik}, the problems of nonholonomic
mechanics are intermixed with those of nonholonomic geometry (the latter
discipline was initiated by Cartan, Wagner and Rashevsky), and this
intermixing has not yet brought any results for the problems of dynamics
itself. In the said papers, for example, the equations of motion are
postulated from the variational principle, invalidity of which for
nonholonomic mechanics was stated as far back as by H.\,Herz. In this
case, one obtains for a disk rolling on a plane the equations of vaconomic
mechanics that do not describe rolling of bodies. (In vaconomic mechanics,
the constraints are realized with the help of added masses; these and
similar issues are discussed in papers~\cite{dirak,kozlov3}.) From the
series of papers in question, only Wagner's paper~\cite{vagner} can,
apparently, be of any interest to a student of dynamics (in this paper,
Wagner gave purely mechanical interpretation of the constraint of Suslov's
problem in contradistinction of the erroneous realization of the same
constraint given by Suslov himself~\cite{suslov}).

Considerably more notable and modern work by
A.\,P.\,Markeev~\cite{markeev1} stands out against the background
of the mentioned monographs and reviews. It was published in 1992, but, unfortunately,
is available only in Russian. Into the next volume we included two of the
most interesting Markeev's papers from the mentioned book, one on the
dynamics of a Woblestone and other on the integrable motion of a ball with a
rotor.

To conclude, let us note that many historical details and new
developments, not included in this essay, are dwelled upon in two our
papers dealing with a rigid body rolling on a fixed surface without sliding,
which is the classical field of application for the results of
nonholonomic dynamics.

We also believe that these two papers are opening a new page in the study
of nonholonomic systems, which is closely connected with the efficient use
of computer-aided methods (systems of analytical calculations,
visualization of motion, numerical experiments). Following this path, one
can obtain results undreamed of in the times of analytical methods alone.
It is in this new direction of research that the major discoveries of the
decades to come will be made.

\end{document}